\documentclass[a4paper,fleqn,usenatbib]{mnras}

\usepackage{graphicx}	% Including figure files
\usepackage{amsmath}	% Advanced maths commands
\usepackage{amssymb}	% Extra maths symbols

\title[Dynamical Roche Lobe]{The Dynamical Roche Lobe in Hierarchical Triples}

\author[R. Di\thinspace Stefano]{
Rosanne Di\thinspace Stefano,$^{1}$\thanks{E-mail: rdistefano@cfa.harvard.edu}
\\
$^{1}$Harvard-Smithsonian Center for  Astrophysics, 60 Garden St., Cambridge MA 02138, US\\
}

\begin{document}
\label{firstpage}
\pagerange{\pageref{firstpage}--\pageref{lastpage}}
\maketitle

\begin{abstract}
The Roche lobe formalism describes mass transfer from one star to another.
We develop an extension to hierarchical triples, considering the case in
which a star donates mass to a companion which is itself a binary. The L1
point moves as the inner binary rotates, and the Roche lobe pulsates with
the period of the inner binary. Signatures of mass transfer may therefore 
be imprinted with the orbital period of the inner binary.  
For some system parameters, the pulsing Roche lobe
 can drive mass transfer at high rates.
Systems undergoing this type of  
mass transfer include those with inner binaries consisting of
compact objects
that will eventually merge, as well as 
progenitors of Type Ia supernovae.    
\end{abstract}
\begin{keywords}
keyword1 -- keyword2 -- keyword3
\end{keywords}

%%%%%%%%%%%%%%%%%%%%%%%%%%%%%%%%%%%%%%%%%%%%%%%%%%

%%%%%%%%%%%%%%%%% BODY OF PAPER %%%%%%%%%%%%%%%%%%
\section{Roche lobe for a Hierarchical Triple}

        The concept of the Roche lobe is important to our 
understanding of binary evolution. 
When a star comes close
to filling its Roche lobe its shape is distorted by tidal and rotational effects. When it
fills its Roche lobe, it can transfer mass directly to its stellar
companion through the L1 point.
For higher order multiples, the simple Roche-lobe picture can serve only as
an approximation \citet{2018arXiv180509338D}. Here we refine the picture in a way that is
useful for the study of  hierarchical triples.  

We begin with the well-studied case of two stars
with masses $M_1$ and $M_2$ in a circular orbit, with the
spin frequency of Star~2 equal to the orbital frequency.
We introduce the triple by replacing Star 1 by two stars,
$a$ and $b$ with the same center of mass as $M_1$ and the
same total mass: $M_{a}+M_{b}=M_1$. The orbital separation between $a$ and
$b$ is small in comparison with the distance between their center of mass
and Star~2. (See Figure~1.)

The L1 point of the original binary is that point lying between 
Star~1 and Star~2 
at which
the centripetal acceleration of a test mass is equal to the
acceleration induced by the combined gravitational pulls 
of Star~1 and Star~2.  
The L1 point is therefore static in the corotating frame.

For a triple, there is generally 
no single frame within which all of the rotational
frequencies are the same.
In addition, the inner and outer orbits may occupy different planes. 
 Because, however, we want to study
cases in which Star~2 experiences tidal deformations and/or transfers
mass, we will employ a ``corotating frame'', coincident with the
plane of the outer orbit,  in which Star~2's spin 
frequency is equal to 
the frequency of the outer orbit. 
The L1 point is that point at which the
centripetal acceleration of a test mass is equal to the
acceleration induced by the combined gravitational pulls of 
Star~{\sl a}, Star~{\sl b}, and
Star~2. Because Star~{\sl a} and Star~{\sl b} are moving, the location of this
equilibrium point also moves.

In \S 2 we explore the motion of L1 points in hierarchical triples.
We take the total mass
of the system to be unity. Thus $M_1+M_2$=1; $M_{a} = f\, M_1$, where
$f$ is a positive number less than unity, and $M_{b}=M_1-M_{a}$.
The distance between $M_1$ (which is at the center of mass of the inner
binary) and $M_2$ is set to unity. The separation $2\, r$ between Star~a
and Star~b must be significantly smaller than unity.  

The energetics of mass moving within the binary system is described by an 
effective potential, $\Phi(r)$, 
which includes both gravitational and rotational effects.  
The L1 point connects two closed equipotential surfaces. 
Each of these equipotential surfaces encloses one of the two stars,
forming that star's Roche lobe.
When Star~2 fills its Roche lobe, 
the L1 point is an escape hatch through which mass can start to travel
toward Star~1. 

The definition of the Roche lobes within a hierarchical triple 
is analogous to the binary-based definition. The difference is that
the L1 point moves, and as it does, the value of the potential at the L1
point also generally changes. At any given instant, the Roche lobe of 
Star 2 is defined by the instantaneous value of $\Phi$ at the L1 point.
As the value of $\Phi$ changes, so does the size and shape of the 
Roche lobe. These same statements apply to the Roche lobe that surrounds the
inner binary.

\begin{figure}
 \includegraphics[width=\columnwidth]{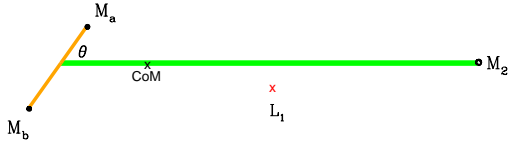}
\vspace{-2.5 in}  
\caption{Hierarchical triple. Star~2 is connected by the thick green line to the
center of mass of the inner binary, whose axis is delineated by an orange
(thinner) line.
The three-body center of mass is marked ``CoM''.  
If the pair $M_a$, $M_b$ were replaced by a single mass $M_1=M_a+M_b$, the
L1 point would be located along the green line. 
The L1 point of the three-body system can, however, 
be located off the axis and it moves as the
binary comprised of $a$ and $b$ rotates.  
}
 \label{fig:rl}
\end{figure}

\begin{figure}
  \includegraphics[width=\columnwidth]{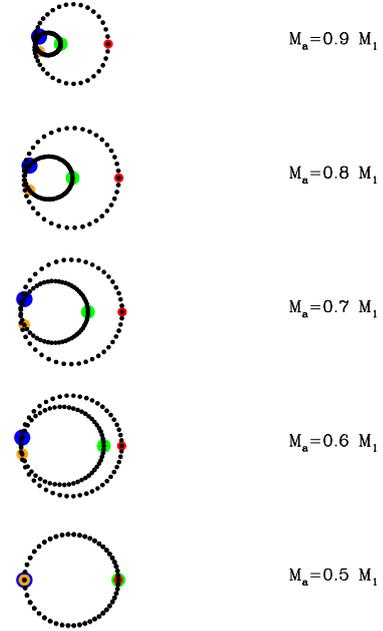}
%\vspace{-4.2 in}  
\caption{{\bf Curves followed by the L1 point.} These triples each have 
$M_1=0.8$. They differ in the value of $M_a$, which varies from $0.9\, M_1$
in the top panel to $0.5\, M_1$ in the bottom panel.  
Each curve displays four special points corresponding to a value of the
phase of the inner orbit: these points that get 
progressively larger and are colored differently, red ($\theta=0$), 
orange ($\theta=\pi/2$), green ($\theta=\pi$), and blue
($\theta=(3\, \pi)/2$), in order of point size. The physical dimensions of the 
diameters of the largest circles shown above are $0.08$ in units where the
separation between star 2 and the center of mass of the inner binary is unity.
As shown in \S 2.5, the size of the curves corresponding to those above would
be smaller for smaller values of the inner-orbit separation.}
 \label{fig:rl}
\end{figure}

\section{Motion of the L1 point and Pulsation of the Roche Lobe}

\subsection{Finding the L1 point}

Star~2 is in a wide orbit with the  
inner binary, 
comprised of Stars {\sl a}  and {\sl b} (Figure~1).
The rotational period of Star~2 is the same as the period of
the outer orbit.
The 
total gravitational force   
on a test mass $dm$ is 
\begin{equation}  
\vec F_{tot} = \vec F_a + \vec F_b + \vec F_2,  
\end{equation} 
where $\vec F_a, \vec F_b$, and $\vec F_2$ are the gravitational forces
exerted by Stars a, b, and 2, respectively.  
In order for the test mass $dm$ to rotate with the outer binary, 
$F_{tot}$ must be equal to the
requisite centripetal force. 
\begin{equation}  
\vec F_{tot} = \vec F_{centripetral}
\end{equation} 
This condition defines the three-dimensional position of the 
L1 point $(X_{L1}, Y_{L1}, Z_{L1})$. 
We take the point $(0, 0, 0)$ to be located at the CoM of the triple.

For the calculations we present here, 
the outer orbit lies in the $x-y$ plane, with Star~2 on the $+x-$axis;
the orbital angular momentum of the outer orbit is
directed along the positive $z-$axis, which points outward toward the reader.  
Thus, in the co-rotating frame, the green axis (Figure~1) is fixed.
The inner binary can rotate in any direction, with $M_a$ and $M_b$
traveling over the surface of three-dimensional spheres of radii
$r_a$  and $r_b$, respectively.     

To create the descriptions and
calculations presented in this paper, we have let the inner and
outer orbital planes coincide. In this case: $x_a=r_a\, cos(\theta)$,\ 
$y_a=r_a\, sin(\theta)$,\ $x_b=r_b\,  cos(\theta+\pi)$,\ $y_b=r_b\, sin(\theta+\pi)$.
The $z$ coordinates of all three masses are zero.
By explicitly considering cases 
in which the inner and outer orbits are aligned, 
we derive results which can be easily described and which can also
be readily generalized to systems in 
which the components of the inner binary move out of 
the plane of the outer orbit.  
 
\subsection{Motion of the L1 point}

The gravitational forces exerted on a test mass  by Stars~{\sl a} 
and {\sl b} change as the 
phase of the inner orbit
changes.  Thus, the point at which Equation~(2) is
satisfied, the L1 point, 
moves as the phase of the inner orbit changes. 
When Stars~{\sl a} and {\sl b} lie 
along the $x-$axis, so that the three stars of the triple 
are aligned,
the magnitude of the
 combined attractive force $\vec F_a+\vec F_b$ 
is generally  larger than 
the magnitude of $\vec F_1$, the force that  would have been
exerted by mass $M_1=M_a+M_b,$ located at the 
center of mass of the inner binary. Thus, to find a new point of equilibrium,
the test mass must slide {\sl away from} the 
inner binary and {\sl toward} Star~2.  
Furthermore, unless $M_a=M_b$, the  position 
of the L1 point is different 
at the inner binary phase $\theta=0$ (when the more massive star, Star~{\sl a}, is
closer to Star~2), from its position at phase $\theta=\pi$ (when the less-massive star,
Star~{\sl b}, is closer to Star~2).  
At phase $\theta=\pi/2,$ the magnitude of the combined force 
$\vec F_a+\vec F_b$ is smaller than for $\theta=0$, or $\theta=\pi$\footnote{
It is also smaller than it would have been for $M_1$ alone.}.
The L1 point therefore moves {\sl toward} the
inner binary and {\sl away from} Star~2.
For other inner-binary phases, 
both Stars~{\sl a} and {\sl b} lie off the $x-$axis, and the 
L1 point must
also move off of the $x-$axis.

Figure~2 illustrates the motion of the L1 point in five different triple-star
systems.  
For each triple in Figure~2, $M_1=M_a+M_b=0.8$ and  
$M_2= 0.2.$  The difference among the triples is in the value of $f$: 
$M_a=f\, M_1,$ as labeled.  
First we consider a single curve, Each point on the curve 
corresponds
to the position of the L1 point at a specific phase of the inner orbit.  
For the purpose  of focusing on the magnitude of the changes,
the origin of the coordinate
system for each of the five  triples shown was chosen to be the symmetry point.
In each curve, the point in red
(smallest of the 4 specially marked points) corresponds to $\theta=0$.
This is the phase at which the combined gravitational pull exerted
by Stars~{\sl a} and {\sl b} is at its maximum. Hence, the L1 point
is farthest from the center of mass of the inner binary and is
closest to Star~2.      
The point in orange (next largest) corresponds to $\theta = \pi/2.$
Proceeding to larger points, 
green is for $\theta= \pi$ and    
blue is for $\theta=(3\, \pi)/2$.
The point in green ($\theta= \pi$) is where all three masses are aligned, and
Star~{\sl b} ($M_b<M_a$) closest to $M_2$.
 
Moving downward from the top curve: as the 
value of $f$ approaches $1/2$,
 the green point moves 
progressively closer to the red point ($\theta=0$).
As long as the values of $M_a$ and $M_b$ are different,
 points for $\theta=\pi/2, ((3\, \pi)/2)$, 
lie slightly below (above) the x-axis, to allow $M_2$ to counter the net
upward (downward) gravitational tug of the inner binary.  
The L1 point executes a closed curve which, except for the
case in which $M_a=M_b$, self intersects. 
On each curve, the time between neighboring black points is $2\, \pi/100.$ 
Thus, larger spatial distances between adjacent black
points show places where the speed of the L1 point was larger,   

In Figure~2, the diameter of the
largest circular curve is $0.08$ of the distance between Star~2 and the
center of mass of the inner binary. 
These dimensions are tied to the separation between the stars in the 
inner orbit. We will explicitly discuss this point iun \S 2.5.

In general, the details of the motion of the L1 point depend on the relative orientations of the 
two orbits, and on whether the orbits are eccentric.  
Depending on the relative orientation
of the planes of the inner and outer orbits, the motion of the L1 point can be
three dimensional. 
The considerations above, of how the combined gravitational force of the 
two components of the inner binary influences the position of the equilibrium point, are
generally valid and provide a way to extend the results to a wide 
range of hierarchical triples.

\subsection{Dependence on $M_1/M_2$}

To systematically explore the effect of varying the mass ratio $M_1/M_2$, we considered
two sets of hierarchical triples.  In the first set (top panel of Figure~3) the masses of the
inner binary were equal ($M_a=M_b$). In the second set (bottom panel),
$M_a=0.8\, M_1$, so that $M_a=4\, M_b$. 
To create the curves in each panel, we varied the value of $M_1$ and $M_2$, with the sum
$M_1+M_2$ always equal to unity.
  The smallest value of $M_1$ we considered
was $0.01,$  which would generally
correspond to a case in which the inner binary consists of orbiting
brown dwarfs or planets. The largest value was $0.99.$   

To  create both the top and bottom panels we  first computed
the position $x(L1,2)$ of the L1 point for a binary with $M_2=1-M_1$. 
The L1 point
lies along the $x-$axis. We then considered configurations
of the inner binary for which the L1 point of the hierarchical triple
also lies along the $x-$axis.

In the top panel of Figure~3, where the masses  
of the components of the inner binary are equal, there are two such
configurations.  
In the first 
(shown in red; topmost curves), Stars~{\sl a} and {\sl b} both lie along the
$x-$axis. In the second configuration
(shown in brown; bottommost curves), 
Stars~{\sl a} and {\sl b} both lie along the
$y-$axis.  

For each configuration of the inner masses, we considered the full range of
values of $M_1$ described above. For each value of $M_1$ we computed the
location $x(L1,3)$ of the L1 point in the corresponding hierarchical triple.
In Figure~3, the difference $x(L1,3)-x(L1,2)$ 
is plotted against $M_1.$ The solid curves
correspond to inner binaries in which the inner-binary separation, 
$R_{in}$ is roughly $0.2$ (see \S 2.5). 
For the dotted curves, $R_{in}$ is only half as large.    

The curves shown in the bottom panel of Figure~3 are analogous, but with 
$M_a=0.8\, M_1$. The inner-binary configurations for which the
L1 point of the hierarchical triple lies along the $x-$ axis are
those in which Stars~{\sl a} and {\sl b} are colinear with Star~2.
There are two independent configurations with this geometry: (1) $M_b$, the
smaller mass, is closer to Star~2; (2) $M_a$ is closer to Star~2. 

The top and bottom panels combine to tell a unified story. When the three
masses are co-aligned, the L1 point moves out, toward Star~2. When the two masses
lie along the perpendicular direction (i.e., the axis proceeding from the
inner-binary CoM to $M_2$ is perpendicular to the axis proceeding from $M_b$ to
$M_a$), the L1 point is pulled back toward the inner binary.    
The magnitude of the change is sensitive to the separation of the inner binary.

We also note that, 
in the range of $M_1$ values over which both $M_1$ and $M_2$ contribute
significantly to the mass budget, the change in L1 depends only slightly 
on the value of $M_1,$ edging up gradually as $M_1$ increases.
At larger values (with $M_1$ larger than $\sim 0.85-0.9$), where Star~{\sl 2} 
could be a brown dwarf
or planet, the motion of the L1 point significantly  increases in magnitude
as $M_1$ increases.

%For the equal-mass inner binary, $R_{in}$ must be smaller than or equal to $0.17$; for the
%hierarchical triple in which the donor provides
%$20\%$ of the total mass, $R_{in}< 0.23.$

\subsection{Pulsation of the Roche Lobe}

When the L1 point moves in toward Star~2, it is moving to places where the 
magnitude, $|\Phi|$, of the gravitational-rotational potential is larger. 
The corresponding equipotential
surface, the Roche lobe, 
is therefore generally {\sl smaller}. On the other hand 
when the L1 point moves away from Star~2, the potential defining the 
Roche lobe has a smaller absolute magnitude and the Roche lobe is
therefore larger.
The result is that the Roche lobe pulsates as the inner orbit proceeds.  

Figure~4 illustrates this effect.  
  In both top and bottom panels, 
the golden curve shows the
Roche lobe of a binary with 
$M_1=0.8$ and $M_2=0.2.$ 
The top panel corresponds to the case in which $M_a=M_b.$ 
The 
inner-orbital phase is $\theta=0,$ (equivalent in this case
to $\theta=\pi$) for the Roche lobe shown in red. As predicted in the
discussion above, the L1 point has moved toward Star~2, i.e., toward
larger values of $|\Phi |$, and the Roche lobe surrounding Star~2 
is smaller than
Star~2's binary-only Roche lobe. The Roche lobe surrounding the inner binary is
stretched
along the horizontal direction.  In brown is the Roche lobe for $\theta=\pi/2$,
for which the lobe surrounding Star~2 is larger, 
and the lobe surrounding the inner
binary is stretched along the vertical direction.   
In both the top and bottom panels, the values of $R_{in}$ are roughly
equal to $0.2$ (\S 2.5). 

In the bottom panel of Figure~4, $M_a=0.8\, M_1$, and $M_b=0.2\, M_1$. 
The curve in blue corresponds to
$\theta=0.,$ in which Star~{\sl a} is closer to Star~2 than is Star~{\sl b}. 
In purple is the
Roche lobe for $\theta=\pi$. In both phases, the total gravitational force
exerted by Stars~{\sl a} and {\sl b} exceeds the force that would 
be exerted by $M_1$, so
that both Roche lobes surrounding Star~2 are smaller 
than for the pure binary case.

\subsection{The Size of the Inner Binary}

The magnitude of the motion of the L1 point 
depends on the size of the inner binary relative to the
separation of the third star from the center of mass of the inner binary. 
If the inner binary is too wide, the triple is not 
hierarchical,
and it also may not be dynamically stable.  

Conditions for dynamical
stability have been derived and also compared with data for hierarchical triples
composed of point masses \citep{1995ApJ...455..640E, 2001MNRAS.321..398M,
2018ApJS..235....6T}. The smallest ratio of orbital periods between the closest orbits
in observed hierarchical triples is typically $\sim 5$, corresponding to
a ratio in orbital separations of $\sim 0.3$.  In the examples considered here,
the largest ratio we considered was  
used $R_{in}\approx 0.2.$\footnote{In the units we employed, with the 
size of the outer orbit set equal to unity, $R_{in}$ is the ratio between the
radii of the inner and outer orbits.}    
The stability conditions may be less restrictive  
when the donor star is filling its Roche lobe, because
tidal forces are also important, and may enhance stability. 
In addition, once mass from the donor begins to interact with the 
inner binary, this interaction may play the key role in determining the fate of the inner binary,
at least until the pool of mass from the donor is exhausted.

The L1-point motions depicted in Figure 2 and in the solid curves of 3 were 
generated by inner binaries
with $R_{in}=0.17$ for the equal-mass inner binary, and $R_{in}=0.23$  
for the inner binary with $M_a=0.8\, M_1.$.  The dotted curves in Figure~3
reveal that there is a significant dependence on the
on $R_{in},$ the radius of the inner binary. 
The dependence is explored in more detail in Figure~5, where we varied the
value of $R_{in}$ .
Here we considered the  

To create Figure 5 we again considered the hierarchical triples with
$M_1=0.8$. For both an equal-mass inner binary (solid lines), and one with
$M_a=0.8\, M_1$ (dashed lines), we varied the value of $R_{in}$, the radius of the
inner orbit, and recorded the values of the $x-$coordinate and $y-$coordinate
of the L1 point and also the radius of the Roche lobe. 
To quantify the change in Roche lobe radius, we defined $\Delta_{RL}$ to  be the
different between the largest and smallest value of $R_L$ achieved during a 
complete cycle of the inner orbit.  Similarly, $\Delta_x$ ($\Delta_y$) 
is the difference between
the largest and smallest value of $x$ ($y$) achieved during one cycle of the inner orbit.  
The figure
shows how the value of each $\Delta$ depends on the value of $R_{in}$. 
For $R_{in} \approx 0.25$ the change in Roche lobe radius over the course of one
inner orbit is roughly $5\%$, and the slope of the log-log plot is 
roughly equal to $-2.$

\begin{figure*}
\centering 
 \includegraphics[width=6.5 in]{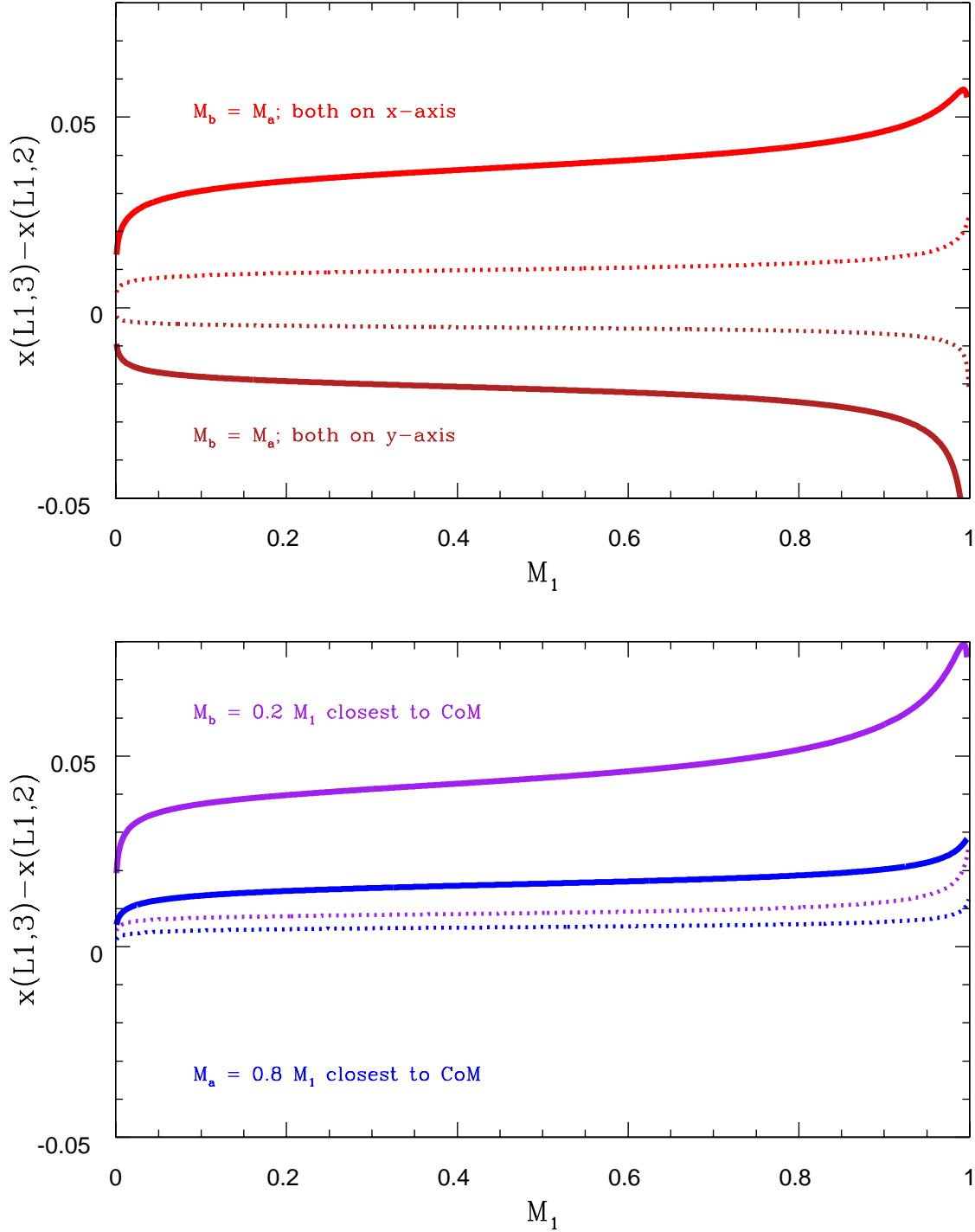}
\caption{
Positions of the L1 point 
for hierarchical triples, as a function of the binary mass, $M_1=M_a+M_b$.
 {\sl Top panel:}  $M_a= M_b =0.5\, M_1$. 
Solid curves: $R_{in}=0.17$; dashed
curves: $R_{in}=0.085.$  
Curves in red (top two curves) 
correspond to both $a$ and $b$ being on the $x$-axis. 
Curves in brown (bottom two curves) 
correspond to both $a$ and $b$ being on the $y$-axis. 
%Here and also in the bottom panels, the inner binary (a-b) 
%has been replaced by a point mass to produce the curves in gold. 
{\sl Bottom panel:}  $M_a= 0.8\, M_1$; Star~$a$ and Star~$b$ 
are both on the x-axis.
Solid curves: $R_{in}=0.23$; dashed
curves: $R_{in}=0.115.$  
Curves in blue (bottom solid curve and bottom dotted curve)  
correspond to the situation in which
$M_a$ 
is closer to Star~2.  
In purple are the curves corresponding to the case in which 
$M_b$ is closer to Star~2.
} 
 \label{fig:rl}
\end{figure*}

\begin{figure*}
\centering 
 \includegraphics[width=6.5 in]{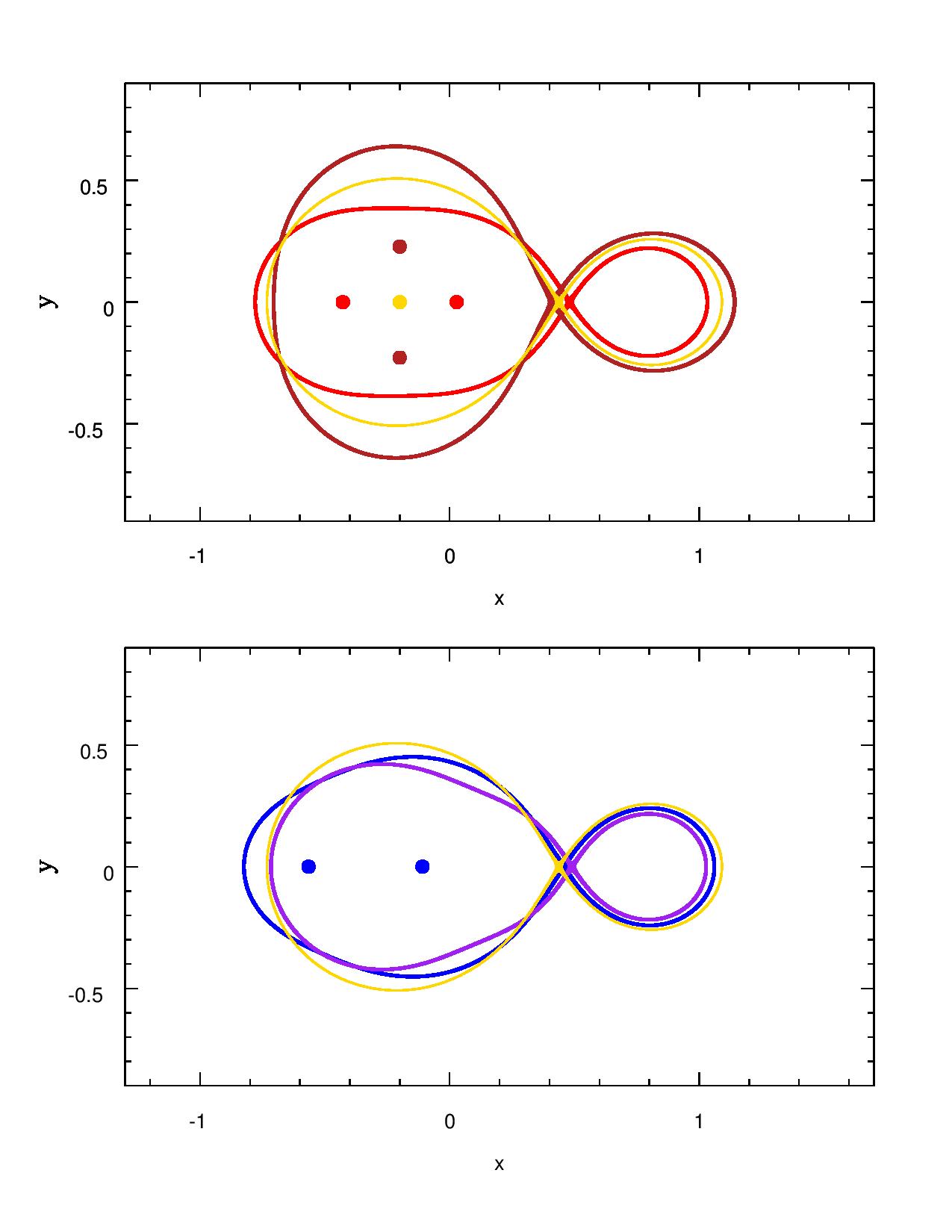}
\caption{
Shapes of the Roche lobes for hierarchical triples. In all cases shown, 
$M_1=0.8.$ {\sl Top:}  $M_a= M_b =0.5\, M_1; R_{in}=0.17$. 
The red curve (which is elongated along the $x-$axis on the left)
 corresponds to both $a$ and $b$ being on the $x$-axis (red points). 
The brown curve (which is elongated along the $y-$axis on the right) 
corresponds to both $a$ and $b$ being on the $y$-axis (brown points). 
Here and also in the bottom panel, the inner binary (a-b) 
has been replaced by a point mass to produce the curve in gold. 
{\sl Bottom:}  $M_a= 0.8\, M_1; R_{in}=0.2$; Star~$a$ and Star~$b$ 
are both on the x-axis.
Curves and points in blue (purple) correspond to the situation in which
$M_a$ ($M_b$) is closer to Star~2. The blue dots mark the inner-star positions
for the blue curve. The positions for the purple curve are not included, simply to 
avoid confusion.  
} 
 \label{fig:rl}
\end{figure*}
\begin{figure*}
\centering 
 \includegraphics[width=6.5 in]{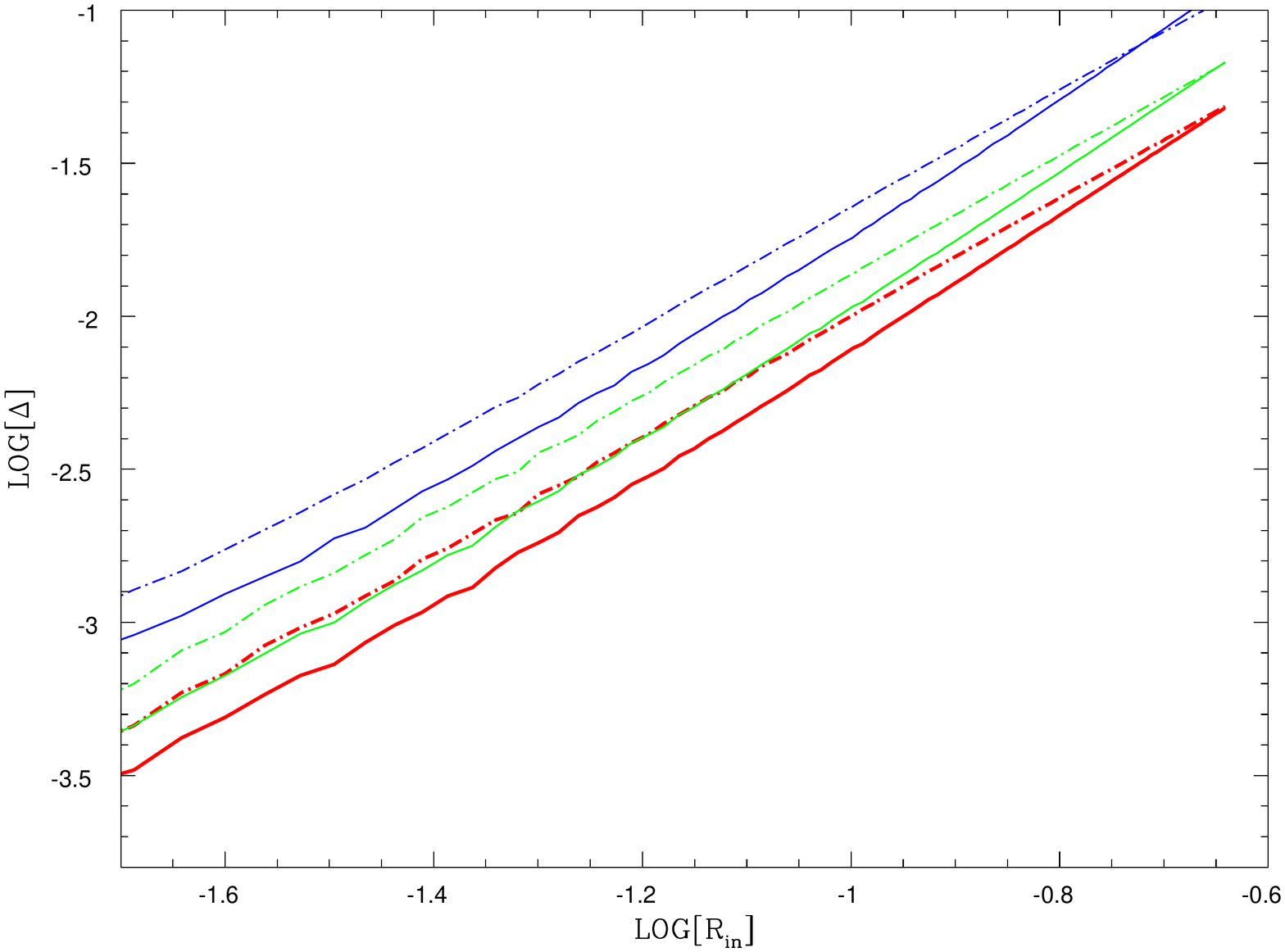}
\caption{The logarithm to the base ten 
of the absolute magnitude of the change in size ($\Delta$) versus the
logarithm to the base ten of $R_{in}$. In red, the change in size refers to
the difference between the maximum and minimum radii achieved by the Roche lobe during
the course of a complete orbit of the inner binary. In green (blue) 
the change in size refers to the absolute magnitude of the maximum difference between the
$x-$coordinate ($y-$coordinate) of the L1 point. In all cases shown here
$M_1=0.8$. Dashed lines
(solid lines) are for hierarchical triples with $M_a=M_b$ ($M_a=4\, M_b$).}      
 \label{fig:rin}
\end{figure*}

% solid: yes_qin_vary3.rin_fin_88; M_1=0.8; M_a=0.8 M_1  
% dash:  yes_qin_vary3.rin_fin_85; M_1=0.8; M_a=0.5 M_1  
% colors are green xdif 
%          blue ydif  
%          red rldif  

\section{Implications}

\subsection{The Dynamical Roche Lobe and New Modes of Mass Transfer}

The donor star in a simple binary comes to fill its Roche lobe 
because the Roche lobe shrinks and/or the donor expands. To 
simplify the discussion below we will discuss the case in which 
the Roche lobe shrinks because 
the separation between the donor and accretor is decreasing
due to the loss of orbital angular momentum associated with, for example, 
magnetic braking. Text describing Roche-lobe filling
due to increasing stellar size  
would be directly analogous.   
When the Roche lobe shrinks to the size of the donor star, the donor
can send mass through the L1 point to its companion. The initiation
of mass transfer influences both the orbital separation and the size of the
donor star. As long as the relative sizes of the Roche lobe and donor star
continue to be well matched,
mass transfer is stable. 

Shifting our consideration to the case in which the accretor is a binary,
we continue to focus on donor-to-accretor mass rations where
stable mass transfer would occur, were the accretor a single star.    
Because the accretor is a binary, the Roche lobe is not only
shrinking, but is also pulsating. Prior to mass transfer, the donor
fits within 
the smallest Roche lobe 
associated with the inner binary.

As time goes on, the average volume of the Roche lobe continues to shrink.
During some portions of the inner orbit, the Roche lobe will touch the 
donor's surface, and then withdraw. 
The relatively small amount of mass lost during the brief
intervals of Roche-lobe filling may not be enough to allow the system to come
into equilibrium. 
Thus, over times longer than the inner orbital period,
the size of the Roche lobe may continue to shrink. When this happens, the
star will overfill the smallest Roche lobe part of the time, 
but may not fill the maximum-radius Roche lobe produced by the inner orbit.
The  
pulsations of the Roche lobe would therefore 
introduce a periodicity to the mass flow.
Preliminary hydrodynamic simulations find that mass loss is enhanced 
with a periodicity equal to that of the inner binary.

\begin{equation}  
\dot M_{dyn} = -\, \kappa\, \frac{M_2}{P_{out}}\Bigg(\frac{R_2-R_{\rm RL}}{R_2}\Bigg)^{n+\frac{3}{2}}, 
\end{equation}  
where $n$ is the polytropic index of Star~2, $P_{out}$ is the orbit period
of the outer orbit, and $\kappa$ is a number whose value is typically
of order unity. [See \citet{2018ApJ...863....5M} and references therein.]   

It is useful to consider some specific examples. Let's
represent $(R_2-R_{RL})/R_2$ by $0.01\, \beta.$ We can consider a  binary with
$M_1=8\, M_\odot$ and take Star~2, with mass $M_2=2\, M_\odot$, to be filling its 
Roche lobe. In our first example, we allow Star~2 to be evolved, 
with $n=3/2$ and a radius of
$50\, R_\odot$. 
 In this case, the orbital period is roughly $94$~days and 
\begin{equation} 
\dot M_{dyn} = 
-\, \kappa\,  \beta^3\, \Bigg(7.8\times 10^{-6}\Bigg)\, \frac{M_\odot}{\rm yr},  
\end{equation} 

To consider a $2\, M_\odot$ main-sequence donor,  we take $n=3.$
The outer binary has an orbital period of $0.76$~days and 
\begin{equation} 
\dot M_{dyn} = 
-\, \kappa\,  \beta^{4.5}\, \Bigg(1\times 10^{-6}\Bigg)\, \frac{M_\odot}{\rm yr}.  
\end{equation}

\subsection{Systems with High $\dot M_{dyn}$}

Equations (4) and (5) show that 
pulsations of the Roche lobe have the potential to produce high values
of the mass transfer rate.
In the case of a hierarchical triple with mass transferred from the outer star,
high rates of accretion
produce high X-ray luminosities if 
one or both of the components of the   
inner binary are compact objects.

When matter is accreting onto a neutron star (NS) or black hole (BH), 
perhaps 
$0.1 \dot M\, c^2$ can be emitted in the form of radiation.
For $\dot M_{dyn} = 10^{-6}M_\odot$~yr$^{-1}$, the associated luminosity is  
$5.7\times 10^{39}$~erg~s$^{-1}$, corresponding to an
ultraluminous X-ray sources (ULX). 
(See \citet{2017Sci...355..817I} and references therein.)  
ULXs are rare, with some galaxies having none, while galaxies with high
rates of star formation may have more than one. 
[See \citet{2013ApJ...778..163B, 2019MNRAS.485.1694D} and references therein.]
 If, therefore, hierarchical mass-transfer binaries 
contribute even one such system per (1-10) galaxies, their
contribution would be significant.

For white-dwarf (WD) accretors,
rates of infall in the ranges consistent with equations (4) and (5) would
produce quasisteady nuclear burning
\citet{Iben.1982, Nomoto.1982, Shen.2007}. The resulting luminosities
would be a few times $10^{38}$~erg~s$^{-1}$. 
It is not clear what fraction of such systems
emit at X-ray wavelengths \citet{rd_dd.2010, rd_sd.2010}, 
although their bolometric luminosities would be
large. Nuclear burning on the surface of a WD can allow infalling mass to
be retained, resulting in a Type~Ia supernova through the so-called
{\sl single degenerate} channel,  
\citet{Rap.1994} or else an accretion-induced collapse \citet{vdHeuvel.1992}.  

Note that $\dot M_{dyn}$ is associated with the pulsation of the Roche lobe.
If the accretor were a single mass (rather than a binary), Roche-lobe filling by the same donor star
would produce mass transfer at some rate $\dot M_0$. The value of $\dot M_0$
is determined by the state of evolution of the donor (e.g., is it expanding?
is it emitting winds?) and on the influence of any dissipative forces that
drive the donor and accretor closer to each other. Thus, the total mass
transfer rate is 
\begin{equation}   
\dot M = \alpha\, \dot M_0 + \beta\, \dot M_{dyn}, 
\end{equation}   
where $\alpha$ and $\beta$ are each constants with values between $0$ and $1$.

Mass transfer to the inner binary can increase the masses of its components.
It can also drive Stars~{\sl a} and {\sl b} closer to each other 
\citet{2018arXiv180509338D}.  Thus, the two stars in the inner binary can
each gain mass. At the same time, the pair is more likely to merge sooner.
This has implications for both the rates of gravitational mergers and for the 
rates of WD-WD mergers, some of which may produce Type~Ia supernovae
through the {\sl double-degenerate} channel.

Whether the components of the inner binary are low-mass objects, stars, or compact objects,
the decrease of the separation between Stars~{\sl a} and {\sl b} 
will decrease the value of $\dot M_{dyn}$, the magnitude of the
Roche lobe's pulsations, and also the size of the L1 point's excursions.
The system will act more like an ordinary binary and, if the inner stars merge,
it will in fact become a binary.

\subsection{Accretion Flow} 

When the accretor is a binary, the L1 point moves around a
small region of the donor star. 
Mass is released from a sequence of different places with different
velocities. Thus, the distances, speeds, and angles of approach to the
accretor's center of mass 
differ from what they would have been in a simple binary. 
Not only is the mass sent from different launch points, but the
potential is also changing. 
The larger the inner binary, the larger the
magnitude of variations in the potential, even
 in the region that would,
for an isolated binary, be occupied by the outer disk.
This may make it impossible for a  circumbinary disk to form.
Accretion within the largest inner binaries may therefore proceed 
through separate accretion by each of its stars.   

For smaller inner binaries, however, circumbinary disks may be expected.
In the case of supermassive BHs, 
``minidisks''  can then form around
the individual components of the close binary 
[e.g., \cite{2017ApJ...838...42B}].  
When conditions are analogous for stellar mass binaries, the same should happen.
Indeed, preliminary
hydrodynamic studies 
\citet{inprep} find that the regions around the
inner stars accumulate matter at relatively high densities.
This discussion indicates that the accretion
flows in hierarchical triples are likely to exhibit a 
variety behaviors and will require careful study.

Apart from the intrinsic interest of the problem, it is important to
understand how much mass can be accreted by each component of the 
inner binary and how much mass is ejected. The amount 
of angular momentum carried away by mass ejected from the vicinity of the
inner binary plays an important role in determining the future of the binary.

\subsection{Do Hierarchical Triples With Mass Transfer Exist?
} 

It is important to demonstrate that there are
three-body systems in which a close binary can come to receive mass
from a star in a wider orbit. In such systems, 
the size of the donor determines the 
size of the Roche lobe: $R_L=R_2$
This, together with the mass ratio, 
$q=M_2/M_1$, sets the size of the outer orbit.  
\begin{equation} 
a_{out}=\frac{R_2}{f(q)}, 
\end{equation} 
where $f(q)=0.49\, q^{\frac{2}{3}}/
(0.6\, q^{\frac{2}{3}} + ln(1+q^{\frac{1}{3}}))$
\citep{1983ApJ...268..368E}.

In addition to the Roche-lobe filling condition above, the system must 
satisfy conditions for orbital stability. Such conditions have been derived
for hierarchical triples in which the components have constant mass
\citet{1995ApJ...455..640E, 2001MNRAS.321..398M}. When
a star that fills or is close to filling its Roche lobe donates mass, the
conditions will be different, as both energy and angular momentum
are associated with tides and with mass flowing through and from
the system. Here we use the 
simplest form of the condition required for orbital stability of the
hierarchical triple,  expressed in terms of a
ratio of the outer to inner orbital periods, $P_{out}/P_{in}> \eta$,
where $\eta = 5\, \eta_0$, where $\eta_0 \approx 1$
 \citet{2001MNRAS.321..398M, 2018AJ....155..160T}.
When tides and/or mass flow serve to stabilize the orbits, the effective
value of $\eta_0$ is smaller.   

 Dynamical stability puts an 
upper limit on the size of the inner binary for a given value of $a_{out}$. 
\begin{equation} 
a_{in}= a_{out}  \Bigg[\frac{M_{in}}{M_T} \, 
\Big(\frac{P_{in}}{P_{out}}\Big)^2\Bigg]^{\frac{1}{3}} 
<\Bigg(\frac{0.34}{\eta_0^\frac{2}{3}}\Bigg)\,  
\Bigg(\frac{R_2}{f(q)}\Bigg)\, \Bigg(\frac{M_{in}}{M_T}\Bigg)^\frac{1}{3} 
\end{equation}
To provide perspective, we note that $f(0.1)=0.21$, and $f(0.9)=0.37$.  
Thus, the separation
between the components of the inner binary can at largest be comparable to the 
radius of the outer Roche-lobe-filling star. 

The condition for orbital stability  must  
be supplemented by the condition that 
mass transfer be stable.  Stability of mass transfer 
generally requires that the donor be 
less massive than the accretor or that, 
if it is more massive, it is able to shrink when it loses mass.   
Putting these conditions together provides
a lower bound on the total mass of the inner binary relative to
the mass of the  main-sequence donor. At the same time, there is an
upper bound on the size of the inner binary.
Fortunately, a wide range 
of hierarchical
triples satisfy these conditions.

\subsubsection{Main-Sequence Donors}

Given the conditions sketched above, it is clear that  
when the donor is on the main sequence,   
$a_{in}$ is generally smaller than a few solar radii.
Stars~{\sl a} and {\sl b} must therefore fit within a small orbit.
One possibility is that both Stars~{\sl a} and {\sl b}
can be stellar remnants.  If the masses of one or both remnants
are comparable to or larger than a solar mass, a gravitational merger
will occur with a Hubble time.   
Mass donated by the outer star can increase their masses,
and shorten the time to merger. In addition, mass transfer may continue through the time of merger, becoming a source of electromagnetic radiation
prior to and during merger. Post merger, mass transfer onto the
merged object may continue. [See \citet{2018arXiv180509338D}  for details.]  
When the donor is a main-sequence star and the inner binary is composed
of stellar remnants, the question of how the triple evolved to its
present state  arise.  This scenario seems most likely in dense clusters.  
Note that, in addition to compact-object inner binaries, there are 
other possibilities. For example, the inner binary could
 contain one or two low-mass stars and/or brown dwarfs.

\subsubsection{Subgiant or Giant Donors}

When the donor is a giant or subgiant, its radius, hence the largest
allowed size of the inner orbit, is larger. This means that inner binaries with
larger donor stars can be accommodated. 
All of the binary types allowed for main-sequence donors are still 
possible. But now 
Stars~{\sl a} and {\sl b} may both be main-sequence stars of larger mass and
therefore with larger radii.  
The case in which the inner binary consists of main sequence stars
is intriguing, because the hierarchical triple could then have a primordial
origin, like many of the higher-order multiples being discovered in ongoing
surveys.  

\subsection{Conclusion}

Mass transfer within hierarchical triples may have 
interesting consequences, and has only begun to be 
systematically explored \citep{2018arXiv180509338D}.
While detailed simulations will play important roles
in elucidating the behavior of such systems,
here we have used the concept of the L1 point and Roche lobe 
to predict and explain key features.   In particular,
over a wide range of ratios  of the inner to outer semimajor axes,
mass transfer will be imprinted with the time signature of
the inner orbital period. Furthermore, for large values of this ratio, 
the pulsation of the Roche lobe may drive mass transfer at high rates.
Mass transfer hierarchical triples with NS and/or
BH accretors may reach the luminosities (above $10^{39}$erg~s$^{-1}$)
associated with ULXs. When the accretor is
a WD dwarf, the rates may be high enough to
promote genuine mass gain by the WD dwarf.  

The 
investigation described here is theoretical, and the results indicate   
that if hierarchical triples are not rare, they may already have been
observed as systems of high astrophysical interest.

\bibliographystyle{mnras}
\bibliography{rl} % if your bibtex file is called example.bib

\begin{thebibliography}{}
\makeatletter
\relax
\def\mn@urlcharsother{\let\do\@makeother \do\$\do\&\do\#\do\^\do\_\do\%\do\~}
\def\mn@doi{\begingroup\mn@urlcharsother \@ifnextchar [ {\mn@doi@}
  {\mn@doi@[]}}
\def\mn@doi@[#1]#2{\def\@tempa{#1}\ifx\@tempa\@empty \href
  {http://dx.doi.org/#2} {doi:#2}\else \href {http://dx.doi.org/#2} {#1}\fi
  \endgroup}
\def\mn@eprint#1#2{\mn@eprint@#1:#2::\@nil}
\def\mn@eprint@arXiv#1{\href {http://arxiv.org/abs/#1} {{\tt arXiv:#1}}}
\def\mn@eprint@dblp#1{\href {http://dblp.uni-trier.de/rec/bibtex/#1.xml}
  {dblp:#1}}
\def\mn@eprint@#1:#2:#3:#4\@nil{\def\@tempa {#1}\def\@tempb {#2}\def\@tempc
  {#3}\ifx \@tempc \@empty \let \@tempc \@tempb \let \@tempb \@tempa \fi \ifx
  \@tempb \@empty \def\@tempb {arXiv}\fi \@ifundefined
  {mn@eprint@\@tempb}{\@tempb:\@tempc}{\expandafter \expandafter \csname
  mn@eprint@\@tempb\endcsname \expandafter{\@tempc}}}

\bibitem[\protect\citeauthoryear{{Bachetti} et~al.,}{{Bachetti}
  et~al.}{2013}]{2013ApJ...778..163B}
{Bachetti} M.,  et~al., 2013, \mn@doi [\apj] {10.1088/0004-637X/778/2/163},
  778, 163

\bibitem[\protect\citeauthoryear{{Bowen}, {Campanelli}, {Krolik}, {Mewes}  \&
  {Noble}}{{Bowen} et~al.}{2017}]{2017ApJ...838...42B}
{Bowen} D.~B.,  {Campanelli} M.,  {Krolik} J.~H.,  {Mewes} V.,   {Noble} S.~C.,
   2017, \mn@doi [\apj] {10.3847/1538-4357/aa63f3}, 838, 42

\bibitem[\protect\citeauthoryear{{Dage}, {Zepf}, {Peacock}, {Bahramian},
  {Noroozi}, {Kundu}  \& {Maccarone}}{{Dage}
  et~al.}{2019}]{2019MNRAS.485.1694D}
{Dage} K.~C.,  {Zepf} S.~E.,  {Peacock} M.~B.,  {Bahramian} A.,  {Noroozi} O.,
  {Kundu} A.,   {Maccarone} T.~J.,  2019, \mn@doi [\mnras]
  {10.1093/mnras/stz479}, 485, 1694

\bibitem[\protect\citeauthoryear{{Di Stefano}}{{Di
  Stefano}}{2010a}]{rd_sd.2010}
{Di Stefano} R.,  2010a, \mn@doi [\apj] {10.1088/0004-637X/712/1/728}, 712, 728

\bibitem[\protect\citeauthoryear{{Di Stefano}}{{Di
  Stefano}}{2010b}]{rd_dd.2010}
{Di Stefano} R.,  2010b, \mn@doi [\apj] {10.1088/0004-637X/719/1/474}, 719, 474

\bibitem[\protect\citeauthoryear{{Di Stefano}}{{Di
  Stefano}}{2018}]{2018arXiv180509338D}
{Di Stefano} R.,  2018, arXiv e-prints

\bibitem[\protect\citeauthoryear{{Eggleton}}{{Eggleton}}{1983}]{1983ApJ...268.%
.368E}
{Eggleton} P.~P.,  1983, \mn@doi [\apj] {10.1086/160960}, 268, 368

\bibitem[\protect\citeauthoryear{{Eggleton} \& {Kiseleva}}{{Eggleton} \&
  {Kiseleva}}{1995}]{1995ApJ...455..640E}
{Eggleton} P.,  {Kiseleva} L.,  1995, \mn@doi [\apj] {10.1086/176611}, 455, 640

\bibitem[\protect\citeauthoryear{{Iben}}{{Iben}}{1982}]{Iben.1982}
{Iben} Jr. I.,  1982, \mn@doi [\apj] {10.1086/160164}, 259, 244

\bibitem[\protect\citeauthoryear{{Israel} et~al.,}{{Israel}
  et~al.}{2017}]{2017Sci...355..817I}
{Israel} G.~L.,  et~al., 2017, \mn@doi [Science] {10.1126/science.aai8635},
  355, 817

\bibitem[\protect\citeauthoryear{{MacLeod}, {Ostriker}  \& {Stone}}{{MacLeod}
  et~al.}{2018}]{2018ApJ...863....5M}
{MacLeod} M.,  {Ostriker} E.~C.,   {Stone} J.~M.,  2018, \mn@doi [\apj]
  {10.3847/1538-4357/aacf08}, 863, 5

\bibitem[\protect\citeauthoryear{{Mardling} \& {Aarseth}}{{Mardling} \&
  {Aarseth}}{2001}]{2001MNRAS.321..398M}
{Mardling} R.~A.,  {Aarseth} S.~J.,  2001, \mn@doi [\mnras]
  {10.1046/j.1365-8711.2001.03974.x}, 321, 398

\bibitem[\protect\citeauthoryear{{Nomoto}}{{Nomoto}}{1982}]{Nomoto.1982}
{Nomoto} K.,  1982, \mn@doi [\apj] {10.1086/159682}, 253, 798

\bibitem[\protect\citeauthoryear{{Rappaport}, {Di Stefano}  \&
  {Smith}}{{Rappaport} et~al.}{1994}]{Rap.1994}
{Rappaport} S.,  {Di Stefano} R.,   {Smith} J.~D.,  1994, \apj, 426, 692

\bibitem[\protect\citeauthoryear{{Shen} \& {Bildsten}}{{Shen} \&
  {Bildsten}}{2007}]{Shen.2007}
{Shen} K.~J.,  {Bildsten} L.,  2007, \mn@doi [\apj] {10.1086/513457}, 660, 1444

\bibitem[\protect\citeauthoryear{{Shr\"oder}, {MacLeod}, {Di\thinspace Stefano}
   \& {Knight}}{{Shr\"oder} et~al.}{2019}]{inprep}
{Shr\"oder} S.,  {MacLeod} M.,  {Di\thinspace Stefano} R.,   {Knight} A.~H.,
  2019

\bibitem[\protect\citeauthoryear{{Tokovinin}}{{Tokovinin}}{2018a}]{2018AJ....1%
55..160T}
{Tokovinin} A.,  2018a, \mn@doi [\aj] {10.3847/1538-3881/aab102}, 155, 160

\bibitem[\protect\citeauthoryear{{Tokovinin}}{{Tokovinin}}{2018b}]{2018ApJS..2%
35....6T}
{Tokovinin} A.,  2018b, \mn@doi [\apjs] {10.3847/1538-4365/aaa1a5}, 235, 6

\bibitem[\protect\citeauthoryear{{van den Heuvel}, {Bhattacharya}, {Nomoto}  \&
  {Rappaport}}{{van den Heuvel} et~al.}{1992}]{vdHeuvel.1992}
{van den Heuvel} E.~P.~J.,  {Bhattacharya} D.,  {Nomoto} K.,   {Rappaport}
  S.~A.,  1992, \aap, 262, 97

\makeatother
\end{thebibliography}

\end{document}